\pdfoutput=1
\documentclass[aps,showpacs,twocolumn,superscriptaddress]{revtex4}

\usepackage{graphicx}
\usepackage{dcolumn}
\usepackage{bm}
\usepackage{color}
\usepackage[normalem]{ulem}
\usepackage[dvipsnames]{xcolor}
\usepackage{hyperref}
\hypersetup{
  colorlinks=true,
  linkcolor=blue,
  citecolor=cyan,
}
\usepackage{mathrsfs}
\usepackage[utf8]{inputenc}
\usepackage{mathtools}
\usepackage{doi}
\usepackage{amsmath}
\usepackage{amssymb}
\usepackage{orcidlink}

\allowdisplaybreaks[3]

\begin{document}

\title{Circular motion and fundamental frequencies of charged particles in dipole magnetic field around magnetized Bocharova--Bronnikov--Melnikov--Bekenstein black holes}

\author{Ozodbek~Abdurakhmonov\orcidlink{0009-0001-0439-6132}} \email{ozodbek992606@gmail.com} 
\affiliation{Institute of Fundamental and Applied Research, National Research University TIIAME, Kori Niyoziy 39, Tashkent 100000, Uzbekistan}

\author{Mardon Abdullaev \orcidlink{0000-0002-0783-7466}}
\email{mardonabdullaev@gmail.com}
\affiliation{Institute of Fundamental and Applied Research, National Research University TIIAME, Kori Niyoziy 39, Tashkent 100000, Uzbekistan}
\affiliation{Institute of Nuclear Physics, Ulugbek 1, Tashkent 100214, Uzbekistan}

\author{Javlon~Rayimbaev~\orcidlink{0000-0001-9293-1838}}
\email{javlon@astrin.uz}
\affiliation{School of Physics, Harbin Institute of Technology, Harbin 150001, People's Republic of China}
\affiliation{Kimyo International University in Tashkent, Shota Rustaveli Street 156, Tashkent 100121, Uzbekistan}

\author{Ilkhomjon~Makhmudov\orcidlink{0000-0002-6026-5292}}
\email{ilhom_makhmudov@mail.ru}
\affiliation{Research Institute of Irrigation and Water Problems, Tashkent, 100187, Uzbekistan}

\author{Saidmuhammad~Ahmedov~\orcidlink{0009-0004-3270-4747}}
\email{s.axmedov@nuu.uz} \affiliation{National University of Uzbekistan, Tashkent 100174, Uzbekistan}

\author{Muhammad~Matyoqubov \orcidlink{0000-0001-8232-6562}} 
\email{m_matyoqubov@mamunedu.uz} \affiliation{Mamun University, Bolkhovuz Street 2, Khiva, Uzbekistan}

\date{\today}

\begin{abstract}
We examine the field produced by a stationary current loop encircling a static Bocharova--Bronnikov--Melnikov--Bekenstein (BBMB) black hole and the response of charged test particles. From the dipole projection of Maxwell's equations, we obtain closed-form interior and exterior radial functions, and we test them against a multipole expansion of the localized source whose radial functions are also elementary. Every horizon-regular mode vanishes at the degenerate horizon $r_+=M$, so the magnetic flux through the horizon is zero. For a loop at $r_0=10M$, the dipole errors in the equatorial potential and polar field remain below $10\%$ for $r\lesssim0.42\,r_0$ and $r\gtrsim3.02\,r_0$, but reach about $40\%$ and $80\%$ close to the loop. Using the resulting potential, we study the effective potential, circular orbits, the innermost stable circular orbit (ISCO), a binding-energy proxy, and epicyclic frequencies. Because the dipole field reverses its equatorial direction across the loop, a positive magnetic coupling $\mathcal{B}$ acts oppositely on the two sides. Above $\mathcal{B}_c\simeq0.0119$, the prograde radially stable orbits split into two bands separated by an unstable gap. At $\mathcal{B}=0.2$, the ISCO moves from $4M$ to $2.97M$, the exterior band begins at $45.46M$, and the ISCO binding-energy proxy grows from $8.14\%$ to $23.39\%$, compared with $5.72\%$ for neutral particles around a Schwarzschild black hole. For negative coupling, exterior orbits can be radially stable but vertically unstable. Higher multipoles preserve the large-coupling trends but shift the near-loop values appreciably, so the near-loop results apply to the projected source rather than to a thin ring.
\end{abstract}
\pacs{04.50.-h, 04.40.Dg, 97.60.Gb}

\maketitle

\section{Introduction}

Scalar fields provide one of the simplest settings in which a black hole might differ from the Kerr--Newman family. For a minimally coupled scalar field, no-hair arguments strongly restrict such a possibility~\cite{Bekenstein1995,HerdeiroRadu2015}. A conformal coupling to curvature changes the situation in a controlled way: the Einstein--conformal scalar equations admit the static BBMB black hole~\cite{Bocharova1970,Bekenstein1974,Bekenstein1975}. The scalar field diverges at the horizon, consistent with the result that static, asymptotically flat, non-extremal black holes with a scalar field bounded on the horizon reduce to Schwarzschild~\cite{Zannias1995}; related restrictions on spherical black holes were discussed in Ref.~\cite{SudarskyZannias1998}. The solution is also unstable under monopole perturbations~\cite{BronnikovKireev1978}, whereas its generalizations with a cosmological constant or a self-interaction~\cite{MTZ2003,AnabalonCisterna2012} show that conformal hair remains a recurring ingredient of exact solutions. The BBMB geometry should therefore be regarded as a useful analytic laboratory rather than as an established astrophysical end state.

This geometry has recently attracted attention as a test bed for strong-gravity phenomena. Studies have addressed its geodesic motion and characteristic radii~\cite{TurimovGeodesic2025}, particle dynamics and acceleration~\cite{TurimovAccel2025}, tidal forces~\cite{TurimovTidal2026}, weak lensing~\cite{Javed2020}, and the quasibound states of scalar fields~\cite{Senjaya2024}. The magnetosphere of a magnetized neutron star has also been analyzed in a BBMB background~\cite{Sayfiyev2025}. More generally, horizon-scale observations of M87* and Sgr~A*~\cite{EHTM87I,EHTSgrAI} have encouraged tests of scalar hair through black-hole shadows~\cite{Khodadi2020}.

Magnetic fields provide another channel for probing such spacetimes. Polarimetric observations indicate organized magnetic structure near the event horizon of M87*~\cite{EHTPolVIII}, while pulsar measurements point to a strong field near Sgr~A*~\cite{Eatough2013}. Electromagnetic processes also play a central role in energy extraction from rotating black holes~\cite{BlandfordZnajek1977}. In the test-field regime, the magnetic field leaves the geometry essentially unchanged, yet it can strongly alter charged-particle motion through the Lorentz force. Since Wald's solution for a black hole in an asymptotically uniform magnetic field~\cite{Wald1974}, this regime has been explored through charged-particle orbits and escape~\cite{FrolovShoom2010,AlZahrani2013,Shiose2014}, quasiharmonic oscillations~\cite{Kolos2015,Tursunov2016}, and chaotic motion~\cite{TakahashiKoyama2009,Kopacek2010}.

An asymptotically uniform field is not the only relevant configuration. Currents localized in the inner part of an accretion flow produce a field that decays at large distances and has different interior and exterior structures. Petterson obtained the field of a current loop around a Schwarzschild black hole~\cite{Petterson1974} and extended the analysis to stationary axisymmetric fields around rotating black holes~\cite{Petterson1975}; Chitre and Vishveshwara also treated the Kerr loop~\cite{ChitreVish1975}. Related studies considered the field lines of a point charge~\cite{HanniRuffini1973}, magnetically induced confinement near a Kerr black hole~\cite{PretiDeFelice2006}, and orbits in multipole fields of Schwarzschild and slowly rotating Kerr spacetimes~\cite{Kerner2025}. Recent work has obtained current-loop fields for the Janis--Newman--Winicour spacetime~\cite{RahmatovJNW2026} and the exponential metric~\cite{TurimovLoop2026}, showing that the radial response depends on the underlying geometry. To the best of our knowledge, the corresponding problem has not been studied for a BBMB black hole.

The BBMB background makes this problem particularly interesting analytically. Its metric function has a double root at the horizon, and the radial functions of the loop become elementary rather than logarithmic or hypergeometric. Moreover, earlier work~\cite{BicakJanis1985} studied magnetic fluxes across black holes, and the degenerate horizon allows direct examination of the flux for each multipole. It therefore remains to be determined how this structure shapes the dipole field, how reliable a dipole truncation is close to the source, and how the resulting field modifies circular orbits, the ISCO, and epicyclic motion.

In this paper, we address these questions for a loop at $r_0=10M$. We first describe the background and the assumptions of the test-field treatment. We next solve the dipole projection of Maxwell's equations, then perform a multipole expansion to assess its accuracy and compute the orthonormal magnetic-field components. Using the resulting potential, we examine the effective potential, circular orbits, the ISCO, and an orbital binding-energy proxy. Finally, we analyze the radial and vertical epicyclic frequencies. Throughout the paper we use units with $G=c=1$ and the metric signature $(-,+,+,+)$.

\section{BBMB black hole spacetime}

The action of Einstein gravity coupled to a conformally invariant scalar field $\Phi$ is
\begin{equation}
    \begin{aligned}
    S={}&\int d^4x\sqrt{-g}\left[\frac{R}{16\pi}-\frac{1}{2}g^{\mu\nu}\Phi_{,\mu}\Phi_{,\nu}\right.\\
    &\left.-\frac{R}{12}\Phi^2\right],
    \end{aligned}
\end{equation}
with the stress-energy tensor of the scalar field
\begin{equation}
    \begin{aligned}
    T_{\mu\nu}={}&\Phi_{,\mu}\Phi_{,\nu}-\frac{1}{2}g_{\mu\nu}g^{\alpha\beta}\Phi_{,\alpha}\Phi_{,\beta}\\
    &+\frac{1}{6}\left(g_{\mu\nu}\Box-\nabla_\mu\nabla_\nu+G_{\mu\nu}\right)\Phi^2 .
    \end{aligned}
\end{equation}
The static BBMB solution~\cite{Bocharova1970,Bekenstein1974} has the line element
\begin{equation}
    ds^2 = -N^2(r)\,dt^2 + \frac{dr^2}{N^2(r)}
    + r^2 \left(d\theta^2 + \sin^2\theta\, d\phi^2\right),
    \label{eq:BBMBmetric}
\end{equation}
with
\begin{equation}
    N^2(r) = \left(1 - \frac{M}{r}\right)^2,
    \label{eq:N2}
\end{equation}
and the scalar field
\begin{equation}
    \Phi(r) = \sqrt{\frac{3}{4\pi}}\;\frac{M}{r - M}.
\end{equation}
The function $N^2$ has a double root at the horizon,
\begin{equation}
    r_+ = M,
\end{equation}
and the Hawking temperature therefore vanishes~\cite{TurimovGeodesic2025}. This metric has the same functional form as that of an extremal Reissner--Nordstr\"om black hole, but the source here is the conformal scalar field rather than an electric charge, and no background electromagnetic field is present. The Ricci scalar vanishes, while the Kretschmann invariant is
\begin{equation}
    R_{\mu\nu\rho\sigma}R^{\mu\nu\rho\sigma}=\frac{8M^2\left(6r^2-12Mr+7M^2\right)}{r^8},
    \label{eq:Kretschmann}
\end{equation}
which remains finite at $r=M$, where $\Phi$ diverges.

\begin{figure}[!htb]
\centering
\includegraphics[width=\columnwidth]{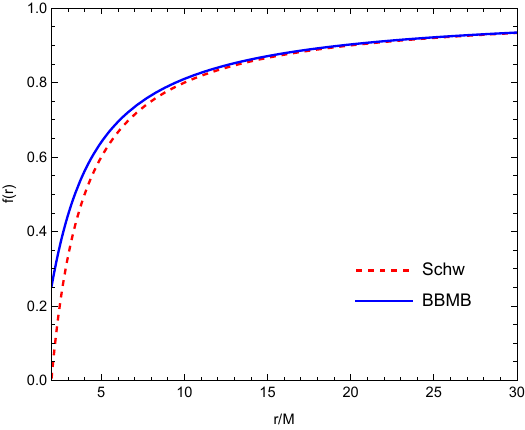}
\caption{Metric function $f(r)$ for the BBMB black hole, $f=N^2=(1-M/r)^2$ (solid blue), and the Schwarzschild black hole, $f=1-2M/r$ (dashed red), as a function of $r/M$.}
\label{fig:f1}
\end{figure}

Figure~\ref{fig:f1} compares $N^2(r)$ with the Schwarzschild function $f(r)=1-2M/r$ at the same mass. Since their difference is $M^2/r^2$, the BBMB curve lies above the Schwarzschild curve at every radius. The separation is largest in the plotted interval at the inner end, where the values at $r=2M$ are $0.25$ and $0$, respectively, and it drops to about $10^{-3}$ at $r=30M$, where $N^2\simeq0.9344$ and $f\simeq0.9333$. Both metrics share the same $1/r$ term, and the conformal scalar enters only through the $M^2/r^2$ contribution, so the two geometries become hard to distinguish in the weak-field region; the difference is already only $0.01$ at $r=10M$. The BBMB horizon lies at $r=M$, below the plotted interval, whereas the Schwarzschild curve reaches zero at $r=2M$. The metric itself therefore differs most strongly in the innermost region.

We treat Eq.~\eqref{eq:BBMBmetric} as a fixed background for the electromagnetic field and for the test particles. This model excludes the instability in Ref.~\cite{BronnikovKireev1978} and field backreaction. The Maxwell action used below contains no direct coupling to $\Phi$, and we assume the particles interact with the scalar field only through the metric. We do not consider models with a direct scalar--particle coupling~\cite{TurimovAccel2025}. The test-field treatment requires the electromagnetic stress-energy to perturb the curvature only slightly. A local condition is $B_{\rm loc}^2L_{\rm curv}^2\ll1$, where $B_{\rm loc}$ is the field measured by a static observer and $L_{\rm curv}$ is the curvature length obtained from Eq.~\eqref{eq:Kretschmann} as the inverse fourth root of the invariant; at the loop radius used below, $L_{\rm curv}\simeq12.66M$. This condition applies away from the idealized wire and differs from the orbital coupling defined in Sec.~\ref{sec:particle}, since a large charge-to-mass ratio can make the Lorentz force significant while the field remains gravitationally negligible.

\section{Magnetic field of a current loop}
\label{sec:maxwell}

\subsection{Dipole solution}

Maxwell's equations in curved spacetime are
\begin{align}
    \nabla_\nu F^{\mu\nu} &= 4\pi J^\mu, \\
    \nabla_\nu \tilde{F}^{\mu\nu} &= 0.
\end{align}
We consider a stationary circular loop of radius $r_0$ in the equatorial plane, carrying a current $I$, with
\begin{equation}
    J^\mu = \frac{I}{r^2 N(r)}\,\delta(r - r_0)\,\delta(\cos\theta)\,(0,0,0,1).
\end{equation}
With this normalization, the coordinate-time current through a meridional half-plane, $\int J^\phi\sqrt{-g}\,dr\,d\theta$, equals $I/N(r_0)$. The loop is an idealized azimuthal current whose radius and strength are input parameters; we do not model the plasma equilibrium that would maintain it. Throughout the numerical analysis, we take $r_0=10M$, which lies outside the neutral ISCO at $4M$ obtained in Sec.~\ref{sec:particle}.

By axisymmetry, the only nonzero component of the vector potential is $A_\phi(r,\theta)$, which satisfies
\begin{equation}
\begin{aligned}
&\frac{1}{\sin^2\theta}
\frac{\partial}{\partial r}
\left[N^2(r)\frac{\partial A_\phi}{\partial r}\right]\\
&\quad+\frac{1}{r^2\sin\theta}
\frac{\partial}{\partial\theta}
\left[\frac{1}{\sin\theta}
\frac{\partial A_\phi}{\partial\theta}\right]\\
&\qquad=
-\frac{4\pi I}{N(r)}
\delta(r-r_0)\delta(\cos\theta).
\end{aligned}
\label{eq:MaxwellAphi}
\end{equation}
Retaining the dipole component of the source, we write $A_\phi(r,\theta)=R(r)\sin^2\theta$. Away from the loop, the radial function obeys
\begin{equation}
    \frac{d}{dr}\left[N^2(r)\frac{dR}{dr}\right] - \frac{2R(r)}{r^2} = 0,
\end{equation}
or, equivalently,
\begin{equation}
    (r-M)^2 R''(r)
+\frac{2M(r-M)}{r}R'(r)
-2R(r)=0.
\end{equation}
Near the horizon, the two solutions behave as $(r-M)$ and $(r-M)^{-2}$. Two linearly independent solutions are
\begin{align}
U(r) &= r^2-M^2, \\
V(r) &= \frac{2r-M}{2(r-M)^2},
\end{align}
which can be verified by direct substitution. Their Wronskian satisfies $N^2(UV'-U'V)=-3$, and at large distances
\begin{equation}
    V(r)=\frac{1}{r}+\frac{3M}{2r^2}+\frac{2M^2}{r^3}+O(r^{-4}).
    \label{eq:Vseries}
\end{equation}
Unlike the Schwarzschild case, no logarithmic term occurs. Regularity at the horizon selects $U$ in the interior, whereas decay at infinity selects $V$ in the exterior. Continuity at $r=r_0$ then gives
\begin{equation}
    A_\phi(r,\theta) = a_0\,F(r)\,\sin^2\theta,
    \label{eq:AphiFinal}
\end{equation}
with
\begin{equation}
F(r)=
\begin{cases}
\displaystyle
(r^2-M^2)\frac{2r_0-M}{2(r_0-M)^2},
& M\leq r\leq r_0,\\[8pt]
\displaystyle
(r_0^2-M^2)\frac{2r-M}{2(r-M)^2},
& r\geq r_0.
\end{cases}
\end{equation}
Projecting Eq.~\eqref{eq:MaxwellAphi} onto the dipole mode and integrating across $r=r_0$ fixes the jump of the radial derivative and yields
\begin{equation}
a_0=\frac{\pi I}{N(r_0)}
=\frac{\pi I}{1-M/r_0}.
\end{equation}
It is convenient to introduce the magnetic dipole moment and the interior field amplitude,
\begin{align}
\mu_* &= a_0(r_0^2-M^2)
      = \frac{\pi I(r_0^2-M^2)}{N(r_0)},\\
B_* &= 2a_0V(r_0)
     = \frac{\pi I(2r_0-M)}
            {N(r_0)(r_0-M)^2}.
\label{eq:Bstar}
\end{align}
For $M\to0$, these expressions reduce to the flat-space dipole moment $\mu_*=\pi I r_0^2$. In the interior region, the potential takes the compact form
\begin{equation}
    A_\phi(r,\theta) = \frac{1}{2}\,B_* \,(r^2 - M^2)\,\sin^2\theta,
    \qquad M\leq r\leq r_0,
    \label{eq:Aphi}
\end{equation}
whereas outside the loop
\begin{equation}
    A_\phi(r,\theta)=\mu_*V(r)\sin^2\theta,
    \qquad r\geq r_0.
    \label{eq:AphiExt}
\end{equation}
The two branches are continuous at $r=r_0$, while $\partial_rA_\phi$ jumps there.

\begin{figure}[!htb]
\centering
\includegraphics[width=\columnwidth]{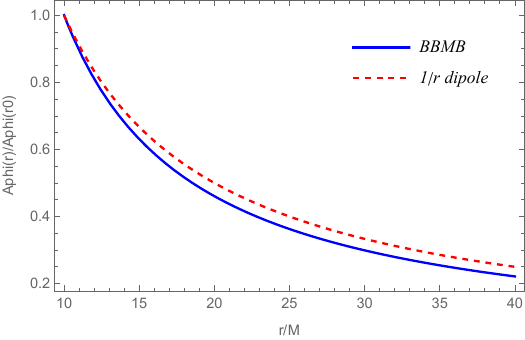}
\caption{Exterior vector potential normalized at the loop, $A_\phi(r,\pi/2)/A_\phi(r_0,\pi/2)$, as a function of $r/M$ for $r_0=10M$ (solid blue), compared with a pure $1/r$ dipole (dashed red).}
\label{fig:Aphi}
\end{figure}

Figure~\ref{fig:Aphi} shows the exterior potential normalized to its value at the loop, together with a pure $1/r$ dipole starting from the same point. The BBMB curve lies below the $1/r$ reference over the whole interval. For example, the two curves give $0.4606$ and $0.5$ at $r=20M$, and $0.2214$ and $0.25$ at $r=40M$. This difference follows from the positive corrections in Eq.~\eqref{eq:Vseries}: $V(r)$ falls more steeply than $1/r$ near the loop, so normalizing at $r_0$ leaves a smaller relative potential at larger radii. Their ratio does not return to unity, but tends to $[r_0V(r_0)]^{-1}\simeq0.853$ for $r_0=10M$. The figure therefore shows that the BBMB geometry changes the effective normalization of the far-field dipole relative to a local value at the loop, while preserving the asymptotic $1/r$ law.

\subsection{Validity of the dipole approximation}
\label{sec:validity}

A thin equatorial loop contains infinitely many odd multipoles, so we must test the retained mode against the localized source. In the BBMB background, the higher modes are also elementary. For the multipole degree $n$, the functions
\begin{equation}
    \begin{aligned}
    U_n(r)&=(r-M)^n\left[r-\frac{r-M}{n+1}\right],\\
    V_n(r)&=(r-M)^{-n-1}\left[r+\frac{r-M}{n}\right]
    \end{aligned}
    \label{eq:UnVn}
\end{equation}
satisfy $(N^2R')'-n(n+1)R/r^2=0$ and $N^2(U_nV_n'-U_n'V_n)=-(2n+1)$; we checked these relations symbolically for $n=1,2,3$, and $5$. With the angular functions $S_n(x)=(1-x^2)P_n'(x)$, $x=\cos\theta$, projection of Eq.~\eqref{eq:MaxwellAphi} gives
\begin{equation}
    \begin{aligned}
    A_\phi^{(N)}(r,\theta)={}&\frac{4\pi I}{N(r_0)}\sum_{\substack{n=1\\ n\ {\rm odd}}}^{N}
    \frac{P_n'(0)}{2n(n+1)}\\
    &\times U_n(r_<)V_n(r_>)\,S_n(\cos\theta),
    \end{aligned}
    \label{eq:Multipole}
\end{equation}
where $r_<=\min(r,r_0)$ and $r_>=\max(r,r_0)$. Since $U_1=U/2$ and $V_1=2V$, the term $n=1$ reproduces Eq.~\eqref{eq:AphiFinal}. Equation~\eqref{eq:UnVn} also shows that $U_n(M)=0$ for every $n\geq1$. Hence the flux through the upper hemisphere of the horizon, $2\pi A_\phi(M,\pi/2)$, vanishes for the dipole mode and for each higher mode. In this sense, the degenerate BBMB horizon expels the axisymmetric flux of the loop, whereas the same flux is nonzero on every equatorial disk bounded by a radius $M<r\leq r_0$.

The two solutions are compared at equal current, mass, and loop radius, using the equatorial relative errors
\begin{equation}
    \begin{aligned}
    \Delta A&=\frac{|A_\phi^{(1)}-A_\phi^{(N)}|}{|A_\phi^{(N)}|},\\
    \Delta B&=\frac{|\partial_rA_\phi^{(1)}-\partial_rA_\phi^{(N)}|}{|\partial_rA_\phi^{(N)}|},
    \end{aligned}
\end{equation}
with $N=1001$. At the equator, $\Delta B$ is also the relative error in $B^{\hat\theta}$. At the radii in Table~\ref{tab:validity}, the results do not change at double precision when $N$ increases from $501$ to $1001$.

\begin{table*}[!t]
\caption{Equatorial errors of the dipole approximation relative to the multipole solution with $N=1001$, for $r_0=10M$. The values are percentages.}
\label{tab:validity}
\begin{ruledtabular}
\begin{tabular}{cccccccc}
$r/r_0$ & 0.20 & 0.30 & 0.45 & 0.90 & 1.10 & 3.00 & 4.50\\
\hline
$100\,\Delta A$ & 0.371 & 1.607 & 5.346 & 39.769 & 41.576 & 3.513 & 1.499\\
$100\,\Delta B$ & 0.952 & 3.835 & 11.968 & 72.936 & 82.493 & 10.149 & 4.416\\
\end{tabular}
\end{ruledtabular}
\end{table*}

For an illustrative tolerance $\Delta A,\Delta B\leq0.1$, the numerical boundaries are $r/r_0\simeq0.4207$ on the interior branch and $r/r_0\simeq3.0216$ on the exterior branch. The approximation is thus reasonably accurate for these equatorial quantities when $r\lesssim0.42\,r_0$ or $r\gtrsim3.02\,r_0$. Close to the loop, however, the errors reach about $40\%$ in the potential and $73$--$82\%$ in the polar field. This is because the higher multipoles carry the wire's singular field, whereas the retained mode corresponds to a current distributed as $\sin\theta$ over the sphere $r=r_0$. Results obtained at $r\approx r_0$ therefore describe the projected source model and should not be read as quantitative predictions for an actual thin ring. Many of the exterior figures below begin at $r=10M$, and this limitation applies directly to their inner parts.

\section{Magnetic field components}

We express the field components in the local orthonormal tetrad carried by a static observer, $u^\mu=N^{-1}(r)\,\delta^\mu{}_t$, who is also a ZAMO because the spacetime does not rotate.

\subsection{Inside the loop}

From Eq.~\eqref{eq:Aphi}, one obtains
\begin{align}
B^{\hat r}(r,\theta)
&= B_*\left(1-\frac{M^2}{r^2}\right)\cos\theta,
\qquad M<r<r_0,\\
B^{\hat\theta}(r,\theta)
&= -B_*N(r)\sin\theta,
\qquad M<r<r_0.
\label{eq:BthetaBBMB}
\end{align}
Both components vanish at the horizon, consistent with the zero flux found above. Far from the horizon, the interior field approaches a uniform field of magnitude $B_*$.

\subsection{Outside the loop}

Using $V'(r)=-r/(r-M)^3$ and Eq.~\eqref{eq:AphiExt}, the exterior components take the closed form
\begin{align}
B^{\hat r}(r,\theta)
&= \frac{\mu_*(2r-M)\cos\theta}{r^2(r-M)^2},
\qquad r>r_0,
\label{eq:BrExt}\\
B^{\hat\theta}(r,\theta)
&= \frac{\mu_*\sin\theta}{r(r-M)^2},
\qquad r>r_0.
\label{eq:BthExt}
\end{align}
For $r\gg r_0,M$, they approach the Newtonian dipole,
\begin{align}
B^{\hat r} &\simeq \frac{2\mu_*\cos\theta}{r^3},\\
B^{\hat\theta} &\simeq \frac{\mu_*\sin\theta}{r^3}.
\end{align}
The polar component changes sign across the loop. At the equator it jumps from $-N(r_0)B_*$ to $\mu_*/[r_0(r_0-M)^2]$, that is, from $-0.9\,B_*$ to about $+0.521\,B_*$ for $r_0=10M$. The radial component remains continuous.

\begin{figure}[!htb]
\centering
\includegraphics[width=\columnwidth]{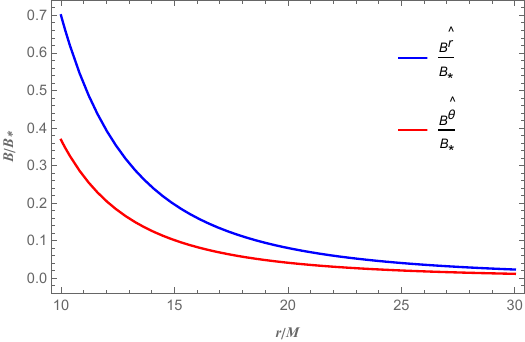}
\caption{Exterior-branch orthonormal magnetic-field components $B^{\hat r}/B_*$ (blue) and $B^{\hat\theta}/B_*$ (red) as functions of $r/M$, for $\theta=\pi/4$ and $r_0=10M$.}
\label{fig:Bofr}
\end{figure}

Figure~\ref{fig:Bofr} shows $B^{\hat r}/B_*$ and $B^{\hat\theta}/B_*$ on the exterior branch at $\theta=\pi/4$. Both components decrease monotonically, with $B^{\hat r}/B_*$ falling from $0.700$ at $r=10M$ to $0.023$ at $r=30M$, and $B^{\hat\theta}/B_*$ from $0.368$ to $0.012$. This fall by a factor of about $30$ is steeper than the factor of $27$ expected from a pure $r^{-3}$ law, because the factors $(r-M)^{-2}$ in Eqs.~\eqref{eq:BrExt} and \eqref{eq:BthExt} add to the decay at finite radius. The ratio of the two components has the simple form
\begin{equation}
    \frac{B^{\hat\theta}}{B^{\hat r}}=\frac{r\tan\theta}{2r-M},
\end{equation}
which explains why the curves look nearly proportional: at $\theta=\pi/4$, the ratio only changes from $0.526$ at $r=10M$ to $0.508$ at $r=30M$, and tends to the Newtonian value $1/2$. Spacetime curvature thus affects the magnitude of the exterior field more strongly than its local direction, with the largest change near the loop, where the accuracy limits of Sec.~\ref{sec:validity} also apply.

\begin{figure}[!htb]
\centering
\includegraphics[width=\columnwidth]{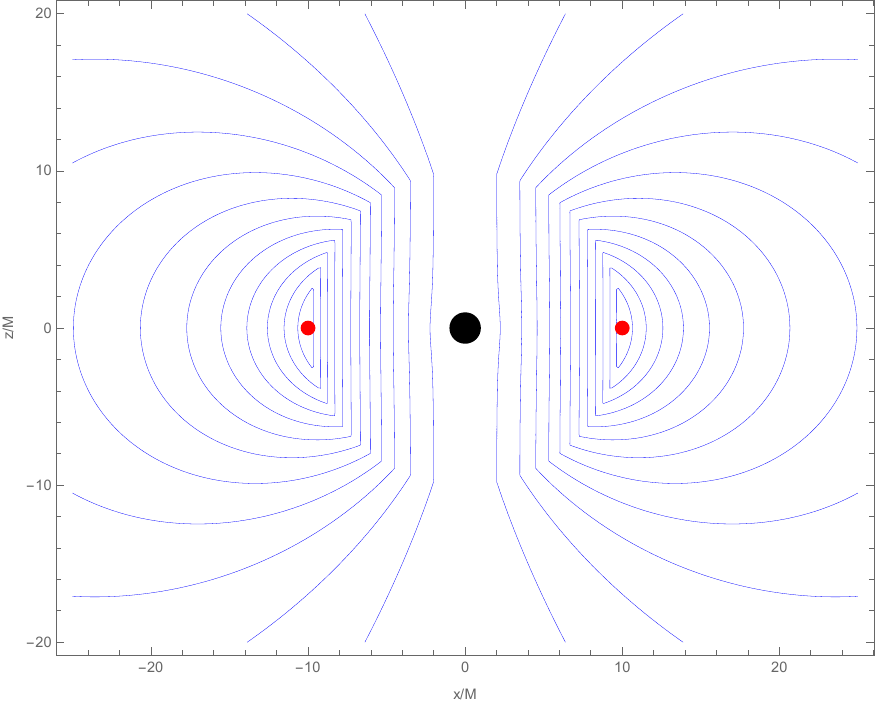}
\caption{Magnetic field lines (contours of $A_\phi$) of the retained dipole mode in the meridional plane for $r_0=10M$. The red dots mark the loop, and the black disk marks the black hole.}
\label{fig:Dipole}
\end{figure}

Figure~\ref{fig:Dipole} presents the field lines, which are the contours of $A_\phi$ in the meridional plane. In the interior, Eq.~\eqref{eq:Aphi} gives $A_\phi=\frac{1}{2}B_*\left(r^2\sin^2\theta-M^2\sin^2\theta\right)$, so that for $r\gg M$ the contours are nearly vertical lines of constant cylindrical radius, corresponding to the almost uniform interior field. They are slightly deformed only near the black hole, and none of the plotted lines end on the horizon, in line with the vanishing flux. Outside the loop, the lines take the closed form of a dipole and encircle the current. The corners at $r=r_0$ are not a resolution effect: they reflect the jump of $B^{\hat\theta}$ across the sphere on which the projected current lies. A complete ring would instead produce closed lines around the wire. This figure therefore clearly shows both the interior--exterior structure of the retained mode and the limit of its validity.

\section{Particle motion}
\label{sec:particle}

\subsection{Effective potential}

We now consider a test particle of mass $m$ and charge $q$, modeled as a structureless body and described by the Hamilton--Jacobi equation
\begin{equation}
g^{\mu\nu}\!\left(\partial_\mu S-qA_\mu\right)\!\left(\partial_\nu S-qA_\nu\right)=-m^2.
\label{eq:HJq}
\end{equation}
Since $A_\phi$ depends on both $r$ and $\theta$, the term $q^2A_\phi^2/(r^2\sin^2\theta)$ couples the radial and angular motion, and Eq.~\eqref{eq:HJq} is not additively separable in these coordinates for $q\neq0$. The equatorial plane is still an invariant surface because of the reflection symmetry, so we set $\theta=\pi/2$ and take
\begin{equation}
    S=m\left(-{\cal E}t+{\cal L}_{\rm tot}\phi\right)+S_r(r),
\end{equation}
where ${\cal E}$ and ${\cal L}_{\rm tot}$ are the specific energy and the specific canonical angular momentum. We define the normalized potential and the dimensionless magnetic coupling parameter as
\begin{equation}
    \psi(r) \equiv \frac{1}{B_*M}\,A_\phi\!\left(r,\tfrac{\pi}{2}\right),
    \qquad
    {\cal B} \equiv \frac{qB_*M}{m},
    \label{eq:calB}
\end{equation}
so that $(q/m)A_\phi(r,\pi/2)={\cal B}\psi(r)$. From Eqs.~\eqref{eq:Aphi} and \eqref{eq:AphiExt},
\begin{equation}
    \psi(r)=
    \begin{cases}
    \displaystyle\frac{r^2-M^2}{2M}, & M\leq r\leq r_0,\\[8pt]
    \displaystyle\frac{(r_0^2-M^2)V(r)}{2MV(r_0)}, & r\geq r_0,
    \end{cases}
    \label{eq:psi}
\end{equation}
which gives $\psi(r_0)=49.5M$ for $r_0=10M$. $\psi$ is continuous at the loop, but $\psi'$ changes sign there: positive inside and negative outside. This feature underlies most of the region-dependent results below. In the following figures, the legend label $q/m$ denotes ${\cal B}$.

The equatorial radial equation and first integrals are
\begin{equation}
    \begin{aligned}
    &\dot r^2={\cal E}^2-V_{\rm eff}(r;{\cal L}_{\rm tot}),\\
    &N^2\dot t={\cal E},
    \qquad
    r^2\dot\phi={\cal L}_{\rm tot}-{\cal B}\psi(r),
    \end{aligned}
\end{equation}
with
\begin{equation}
\begin{aligned}
&V_{\rm eff}(r;\,{\cal L}_{\rm tot})\\
&\quad=N^2(r)\left[1+\frac{1}{r^2}\bigl({\cal L}_{\rm tot}-{\cal B}\,\psi(r)\bigr)^2\right].
\end{aligned}
\label{eq:VeffChargedBBMB}
\end{equation}
The potential depends on the mechanical angular momentum $K={\cal L}_{\rm tot}-{\cal B}\psi$, rather than on ${\cal L}_{\rm tot}$ alone. For ${\cal B}=0$, it reduces to the neutral form
\begin{equation}
    V_{\rm eff}(r) = \left(1 - \frac{M}{r}\right)^2 \left(1 + \frac{{\cal L}^2}{r^2}\right).
    \label{eq:Veff}
\end{equation}

\begin{figure*}[!t]
\centering
\includegraphics[width=0.48\textwidth]{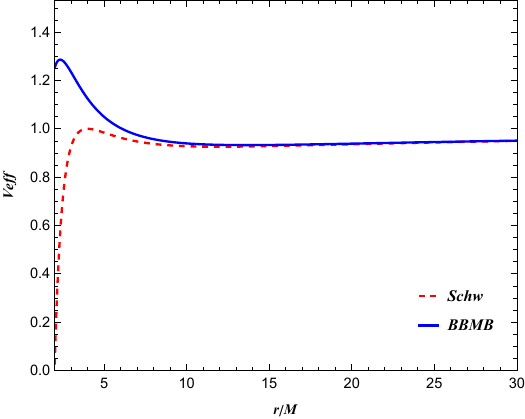}
\includegraphics[width=0.48\textwidth]{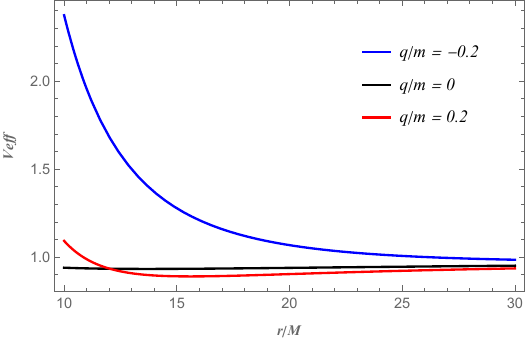}
\caption{Effective potential $V_{\rm eff}$ as a function of $r/M$ for ${\cal L}_{\rm tot}=4M$. Left panel: neutral particles around the BBMB (solid blue) and Schwarzschild (dashed red) black holes. Right panel: particles in the BBMB spacetime on the exterior branch for $r_0=10M$, with ${\cal B}=-0.2$ (blue), $0$ (black), and $0.2$ (red).}
\label{fig:Veff}
\end{figure*}

Figure~\ref{fig:Veff} shows the effective potential at fixed ${\cal L}_{\rm tot}=4M$. In the left panel, the BBMB potential has a maximum of $1.2862$ at $r=2.342M$ and a minimum of $0.9326$ at $r=13.656M$. The Schwarzschild potential has a lower barrier of $1$ at $r=4M$ and a minimum of $0.9259$ at $r=12M$. The BBMB curve lies above the Schwarzschild curve, and the two practically merge at large distances, giving $0.9511$ and $0.9499$ at $r=30M$. The higher and more inward barrier follows from the extra positive $M^2/r^2$ term in $N^2$. For the same angular momentum, a neutral particle therefore meets a stronger centrifugal barrier around the BBMB black hole, and a stable minimum lies slightly farther out.

The right panel shows the exterior magnetic effect at the same canonical angular momentum. At $r=10M$, $V_{\rm eff}$ equals $2.375$, $0.940$, and $1.092$ for ${\cal B}=-0.2$, $0$, and $0.2$, respectively. This ordering is not monotonic in ${\cal B}$, because the potential depends on $K^2$. At the loop, ${\cal B}\psi=\pm9.9M$, which gives $K=13.9M$ for ${\cal B}=-0.2$ and $K=-5.9M$ for ${\cal B}=0.2$. The second case therefore corresponds to retrograde mechanical motion despite the positive canonical angular momentum. For ${\cal B}=0.2$, the potential crosses the neutral curve at $r\simeq12.03M$, where $K=-{\cal L}_{\rm tot}$, and it has a minimum of $0.890$ at $r\simeq15.61M$. This minimum belongs to the negative-$K$ family and should not be confused with the prograde orbits discussed below. For ${\cal B}=-0.2$, the curve decreases throughout the plotted interval, and its minimum lies outside it, at $r\simeq45.2M$. All curves approach unity at large distances, where $\psi\to0$ and the magnetic interaction disappears. Hence, the sign of the coupling cannot by itself determine whether a barrier grows or falls, since this depends on where $K$ lies relative to zero.

\subsection{Circular orbits}

Circular orbits satisfy
\begin{equation}
V_{\rm eff}(r;{\cal L}_{\rm tot})={\cal E}^2,
\qquad
\left.\frac{\partial V_{\rm eff}(r;{\cal L}_{\rm tot})}
{\partial r}\right|_{{\cal L}_{\rm tot}}=0.
\end{equation}
Introducing the auxiliary quantity
\begin{equation}
\label{eq:LtotDef}
y\equiv \frac{{\cal L}_{\rm tot}-{\cal B}\,\psi(r)}{r},
\end{equation}
which is proportional to the mechanical angular momentum; the condition $\partial_rV_{\rm eff}=0$ gives a quadratic equation,
\begin{equation}
\label{eq:quad}
\begin{aligned}
&\bigl[r\,N^{2}{}'(r) - 2N^{2}(r)\bigr]\,y^2\\
&\quad-2\,N^{2}(r)\,{\cal B}\,\psi'(r)\,y
+r\,N^{2}{}'(r)=0,
\end{aligned}
\end{equation}
with
\begin{equation}
\label{eq:Nprime}
N^2{}'(r)=\frac{2M}{r^2}-\frac{2M^2}{r^3}.
\end{equation}
Choosing the prograde (larger) root, the conserved quantities and orbital frequency are
\begin{equation}
\label{eq:ELOmega}
\begin{aligned}
{\cal E}(r)&=\sqrt{\,N^2(r)\bigl(1+y^2\bigr)} ,\\
\Omega(r)&=\frac{d\phi}{dt}=\frac{N^2(r)\,y}{r\,{\cal E}(r)},\\
{\cal L}_{\rm tot}(r)&=r\,y+{\cal B}\,\psi(r).
\end{aligned}
\end{equation}
Since the bracket in Eq.~\eqref{eq:quad} is proportional to $-(r-M)(r-2M)$, the neutral orbits exist for $r>2M$. For ${\cal B}=0$, one finds
\begin{align}
{\cal E}^2(r) &= \frac{(r-M)^3}
{r^2\bigl(r-2M\bigr)},
\label{eq:Ecirc}\\
{\cal L}^2(r) &= \frac{Mr^2}{r-2M},
\label{eq:Lcirc}\\
\Omega_K(r)&=\frac{\sqrt{M(r-M)}}{r^2}.
\label{eq:OmegaK_BBMB}
\end{align}
These expressions yield the characteristic radii listed in Table~\ref{tab:neutral}; Ref.~\cite {TurimovGeodesic2025} also analyzes these radii for this geometry. They all lie inside their Schwarzschild counterparts, for which we use the standard values~\cite{BPT1972}.

\begin{table*}[!t]
\caption{Characteristic neutral-particle quantities in the BBMB and Schwarzschild spacetimes. The final column gives the ISCO binding-energy proxy $\eta$ of Eq.~\eqref{eq:eta} as a percentage.}
\label{tab:neutral}
\begin{ruledtabular}
\begin{tabular}{ccccccc}
Spacetime & $r_{\rm ph}/M$ & $r_{\rm mb}/M$ & $r_{\rm ISCO}/M$ & ${\cal L}_{\rm ISCO}/M$ & ${\cal E}_{\rm ISCO}$ & $100\,\eta$\\
\hline
BBMB & 2 & $(3+\sqrt{5})/2\simeq2.618$ & 4 & $2\sqrt{2}\simeq2.828$ & $3\sqrt{6}/8\simeq0.9186$ & 8.144\\
Schwarzschild & 3 & 4 & 6 & $2\sqrt{3}\simeq3.464$ & $\sqrt{8/9}\simeq0.9428$ & 5.719\\
\end{tabular}
\end{ruledtabular}
\end{table*}

\begin{figure}[!htb]
\centering
\includegraphics[width=\columnwidth]{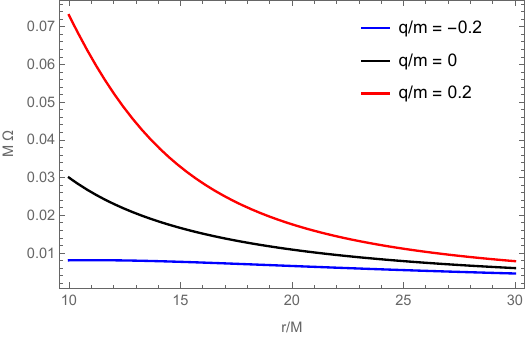}
\caption{Orbital frequency $M\Omega$ of prograde circular orbits on the exterior branch as a function of $r/M$ for $r_0=10M$, with ${\cal B}=-0.2$ (blue), $0$ (black), and $0.2$ (red).}
\label{fig:Momega}
\end{figure}

Figure~\ref{fig:Momega} shows the orbital frequency $M\Omega$ on the exterior branch. At $r=10M$, the frequencies are $0.00805$, $0.0300$, and $0.0731$ for ${\cal B}=-0.2$, $0$, and $0.2$, respectively, and the neutral curve agrees with Eq.~\eqref{eq:OmegaK_BBMB}. The ordering has a direct physical origin. For $qB_*>0$, the exterior equatorial field gives a prograde particle an inward Lorentz force, so a larger azimuthal velocity is required to balance it, and the frequency increases. For $qB_*<0$, the force points outward and partly supports the orbit, so the required velocity falls. The ${\cal B}=-0.2$ curve is correspondingly almost flat, with a broad maximum of about $0.0081$ near $r\simeq11M$. At $r=10M$, these effects are strong, as shown by the values $y=0.090$, $0.354$, and $1.392$ for the three couplings. The separation decreases with distance as $\psi'$ decays, and at $r=30M$ the frequencies are $0.00455$, $0.00598$, and $0.00782$. Thus, in the exterior region, the sign of the coupling clearly imprints itself on the orbital frequency near the loop, while Keplerian behavior gradually returns at larger radii. Because the sign of $\psi'$ reverses inside the loop, the same coupling produces the opposite force direction in the interior, as discussed next.

\subsection{ISCO and stable bands}

Circular orbits are radially stable when $\partial_r^2V_{\rm eff}>0$, and the edges of the stable regions are determined by
\begin{equation}
\partial_r V_{\rm eff}=\partial_r^2 V_{\rm eff}=0.
\label{eq:ISCOcond}
\end{equation}
We take the derivatives at fixed ${\cal L}_{\rm tot}$. Because $\psi'$ jumps at the loop, we impose these conditions separately on the two branches of Eq.~\eqref{eq:psi}, and exclude $r=r_0$ for ${\cal B}\neq0$. We denote $r_{\rm ISCO}$ as the innermost radially stable prograde orbit, and $r_{\rm ext}$ as the inner edge of the stable exterior band when this edge lies above the loop. For ${\cal B}=0$, the electromagnetic interaction disappears and the two branches give the same neutral orbits. The neutral ISCO values $r_{\rm ISCO}=4M$, ${\cal L}=2\sqrt{2}M$, and ${\cal E}=3\sqrt{6}/8$ provide the reference for the charged case.

\begin{figure*}[!t]
\centering
\includegraphics[width=1\textwidth]{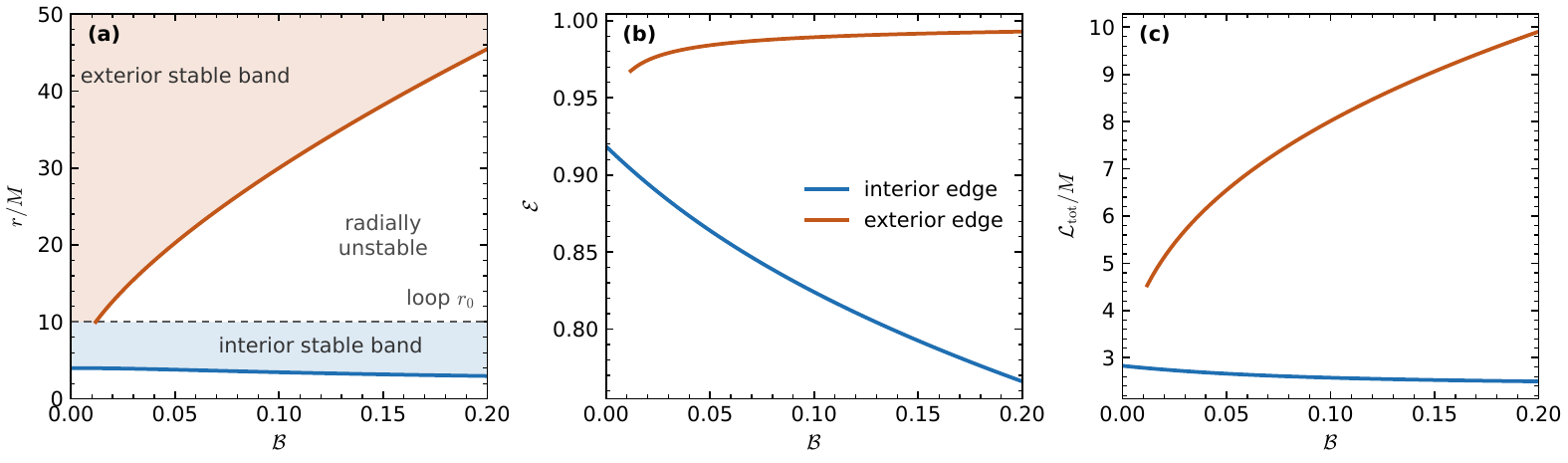}
\caption{Prograde circular orbits for $r_0=10M$ as functions of ${\cal B}$. (a) Radially stable bands, with the interior edge $r_{\rm ISCO}$ (blue) and the exterior edge $r_{\rm ext}$ (orange), which exists for ${\cal B}>{\cal B}_c\simeq0.0119$; the dashed line marks the loop. (b) Specific energy and (c) specific canonical angular momentum at these edges.}
\label{fig:Risco}
\end{figure*}

Figure~\ref{fig:Risco} shows the stable bands and associated conserved quantities for ${\cal B}\geq0$, and Table~\ref{tab:isco} lists the corresponding values. In panel (a), the interior edge $r_{\rm ISCO}$ decreases with the coupling, from $4M$ at ${\cal B}=0$ to $3.96M$ at ${\cal B}=0.02$ and $2.97M$ at ${\cal B}=0.2$. For $qB_*>0$, the interior Lorentz force on a prograde particle points outward and partly supports the orbit against gravity, which is consistent with this inward shift, although the force direction alone does not fix it, as the negative-coupling case below shows. The exterior behavior is very different. For ${\cal B}$ up to the critical value ${\cal B}_c\simeq0.0119$, the orbits immediately outside the loop remain radially stable. Above ${\cal B}_c$, they lose stability, and a radially unstable gap $r_0<r<r_{\rm ext}$ opens between the two stable bands. The edge $r_{\rm ext}$ then grows from $12.66M$ at ${\cal B}=0.02$ to $45.46M$ at ${\cal B}=0.2$. This value is the innermost stable orbit outside the loop, rather than the ISCO of the full configuration.

Panels (b) and (c) show that the two edges also differ energetically. At the interior edge, ${\cal E}$ falls from $0.9186$ to $0.7661$ and ${\cal L}_{\rm tot}$ falls from $2.828M$ to $2.498M$ over $0\leq{\cal B}\leq0.2$, reflecting a smaller radius and a smaller mechanical angular momentum, with $y$ dropping from $0.707$ to $0.577$. In contrast, at the exterior edge the energy rises towards unity, reaching $0.9930$, while ${\cal L}_{\rm tot}$ increases to $9.905M$. Since the exterior edge moves to larger radii, where the gravitational binding is weaker, this energy increase is expected; the larger canonical angular momentum also includes the electromagnetic contribution ${\cal B}\psi$. The piecewise character of the source thus produces a clear separation between a strongly bound interior band and a weakly bound exterior band, directly tied to the localized source.

\begin{table*}[!t]
\caption{Prograde radial marginal-stability points of the retained dipole model for $r_0=10M$, with the interior ISCO and the exterior band edge. Dashes indicate that no gap exists. The percentages give the binding-energy proxy $\eta$ of Eq.~\eqref{eq:eta}.}
\label{tab:isco}
\begin{ruledtabular}
\begin{tabular}{ccccccccc}
${\cal B}$ & $r_{\rm ISCO}/M$ & ${\cal E}_{\rm ISCO}$ & ${\cal L}_{\rm ISCO}/M$ & $100\,\eta$ & $r_{\rm ext}/M$ & ${\cal E}_{\rm ext}$ & ${\cal L}_{\rm ext}/M$ & $100\,\eta_{\rm ext}$\\
\hline
$-0.20$ & 3.3350 & 1.33712 & 4.4139 & $-33.712$ & -- & -- & -- & --\\
$-0.10$ & 3.6109 & 1.09273 & 3.4896 & $-9.273$ & -- & -- & -- & --\\
$-0.05$ & 3.8365 & 0.99507 & 3.1128 & 0.493 & -- & -- & -- & --\\
$-0.02$ & 3.9655 & 0.94640 & 2.9285 & 5.360 & -- & -- & -- & --\\
$0$ & 4.0000 & 0.91856 & 2.8284 & 8.144 & -- & -- & -- & --\\
$0.02$ & 3.9615 & 0.89436 & 2.7485 & 10.564 & 12.6596 & 0.97439 & 5.1272 & 2.561\\
$0.05$ & 3.7954 & 0.86391 & 2.6619 & 13.609 & 20.2595 & 0.98411 & 6.5476 & 1.589\\
$0.10$ & 3.4604 & 0.82404 & 2.5760 & 17.596 & 30.0047 & 0.98932 & 8.0135 & 1.068\\
$0.15$ & 3.1844 & 0.79249 & 2.5277 & 20.751 & 38.1717 & 0.99163 & 9.0624 & 0.837\\
$0.20$ & 2.9711 & 0.76607 & 2.4982 & 23.393 & 45.4588 & 0.99298 & 9.9054 & 0.702\\
\end{tabular}
\end{ruledtabular}
\end{table*}

Negative coupling, not shown in Fig.~\ref{fig:Risco}, is listed in Table~\ref{tab:isco}. In this case, no exterior radial gap appears over the tested interval $-0.2\leq{\cal B}<0$. The interior ISCO still moves inward, to $3.61M$ at ${\cal B}=-0.1$ and $3.34M$ at ${\cal B}=-0.2$, even though the interior Lorentz force now points towards the black hole. The ISCO shift therefore does not follow from the sign of the force alone: marginal stability also depends on $\psi''$ and on the radial variation of the force. The specific energy behaves differently, since it increases as ${\cal B}$ decreases, reaching ${\cal E}_{\rm ISCO}=1$ at ${\cal B}\simeq-0.0528$.

To test these results, we repeated the calculation with the multipole potential of Eq.~\eqref{eq:Multipole}, using $\psi^{(N)}=A_\phi^{(N)}(r,\pi/2)/(B_*M)$ with $N=1001$ and with $B_*$ still given by Eq.~\eqref{eq:Bstar}. Table~\ref{tab:isco_multipole} compares the outcomes. The large-coupling trends survive: for ${\cal B}\geq0.05$ the ISCO moves inward, and the outer stability boundary moves outward with the coupling. Quantitatively, the outer boundary differs by about $4\%$ at ${\cal B}=0.2$, where it is far from the loop, but by about $38\%$ at ${\cal B}=0.02$. The small inward shift of the dipole ISCO at ${\cal B}=0.02$ is not robust either, since the multipole sum gives $4.05M$, slightly above the neutral value. The dipole model therefore captures the overall behavior at larger couplings, whereas ${\cal B}_c$ and the near-loop values are properties of the projected source. We did not map the stability structure immediately around the wire, which would require a finite source cross-section.

\begin{table*}[!t]
\caption{Comparison of the dipole model with the multipole sum of Eq.~\eqref{eq:Multipole} with $N=1001$, at the same current and loop radius $r_0=10M$. We define ${\cal B}$ using the dipole amplitude $B_*$ in both cases. The exterior edge is the outer radial-stability boundary.}
\label{tab:isco_multipole}
\begin{ruledtabular}
\begin{tabular}{ccccc}
${\cal B}$ & $r_{\rm ISCO}/M$ (dipole) & $r_{\rm ISCO}/M$ (multipole) & $r_{\rm ext}/M$ (dipole) & $r_{\rm ext}/M$ (multipole)\\
\hline
$-0.20$ & 3.335 & 3.218 & -- & --\\
$-0.10$ & 3.611 & 3.465 & -- & --\\
$0.02$ & 3.962 & 4.048 & 12.660 & 17.463\\
$0.05$ & 3.795 & 3.905 & 20.260 & 23.928\\
$0.10$ & 3.460 & 3.513 & 30.005 & 32.760\\
$0.20$ & 2.971 & 2.980 & 45.459 & 47.404\\
\end{tabular}
\end{ruledtabular}
\end{table*}

\subsection{ISCO binding-energy proxy}

As a simple measure of orbital binding, we define the energy per unit rest mass that a particle must shed to pass from rest at infinity to the ISCO,
\begin{equation}
    \eta=1-{\cal E}\big|_{r=r_{\rm ISCO}}.
    \label{eq:eta}
\end{equation}
For a neutral particle, $\eta=8.144\%$ for the BBMB black hole, which exceeds the Schwarzschild value of $5.719\%$. At ${\cal B}=0.2$, the interior proxy reaches $23.39\%$, whereas the proxy at the exterior edge falls to $0.70\%$. For negative coupling, $\eta$ falls, and it changes sign at ${\cal B}\simeq-0.0528$. A negative value means a particle starting from rest at infinity cannot settle on that orbit by losing energy alone; it does not, by itself, imply that the circular orbit is unstable.

We use $\eta$ only as an orbital diagnostic. Our equations describe conservative motion in prescribed fields and do not include radiation or disk physics. Identifying $\eta$ with a radiative efficiency would require additional assumptions used in thin-disk models~\cite{PageThorne1974}, including an inner edge at the ISCO and negligible torque there. Moreover, the ISCO energy shift in the present problem arises from a prescribed field whose source is not modeled self-consistently. The larger proxy at positive coupling therefore shows how the ISCO energy responds to the field, but it does not establish a higher luminosity of a magnetized accretion flow.

\section{Fundamental epicyclic frequencies}\label{sec:epicyclic}

The charged-particle equation of motion is
\begin{equation}
\begin{aligned}
&\frac{d u^\alpha}{d\lambda}+\Gamma^{\alpha}_{\mu\nu}u^\mu u^\nu=\frac{q}{m}\,F^\alpha{}_\beta u^\beta,\\
&u^\mu u_\mu=-1.
\end{aligned}
\label{eq:nongeo}
\end{equation}
We perturb a circular equatorial orbit, for which $u^\mu=(u^t,0,0,\Omega u^t)$, by setting
\begin{equation}
   r(\lambda)=r_{\rm c}+\delta r(\lambda),\qquad
\theta(\lambda)=\frac{\pi}{2}+\delta\theta(\lambda),
\end{equation}
and linearize in $\delta r$ and $\delta\theta$. Using the off-equatorial dependence $A_\phi=B_*M\psi(r)\sin^2\theta$, the radial and vertical epicyclic frequencies measured in coordinate time are
\begin{align}
\kappa_r^2(r_{\rm c})
&=
\left.
\frac{1}{2(u^t)^2}
\frac{\partial^2 V_{\rm eff}(r;{\cal L}_{\rm tot})}{\partial r^2}
\right|_{\substack{r=r_{\rm c}\\{\cal L}_{\rm tot}=\mathrm{const}}},
\label{eq:kappaR}\\
\kappa_\theta^2(r_{\rm c})
&=
\frac{
\bigl[{\cal L}_{\rm tot}-{\cal B}\psi(r_{\rm c})\bigr]
\bigl[{\cal L}_{\rm tot}+{\cal B}\psi(r_{\rm c})\bigr]
}{
r_{\rm c}^{4}(u^t)^2
},
\label{eq:kappaTh}
\end{align}
with $u^t={\cal E}/N^2$. For ${\cal B}=0$, the vertical frequency equals the Keplerian frequency owing to spherical symmetry, while the radial frequency has the closed form
\begin{equation}
    \begin{aligned}
    \kappa_r^2&=\frac{M(r-4M)(r-M)^2}{r^6}\\
    &=\Omega_K^2\left(1-\frac{4M}{r}\right)\left(1-\frac{M}{r}\right).
    \end{aligned}
    \label{eq:kappaRneutral}
\end{equation}
At large distances, $\kappa_r^2/\Omega_K^2\simeq1-5M/r$, compared with $1-6M/r$ in Schwarzschild spacetime. Radial and vertical frequencies of this kind enter orbital models of quasi-periodic oscillations~\cite{StellaVietri1999,RemillardMcClintock2006}, and Refs.~\cite{Kolos2015,Tursunov2016} studied their charged-particle counterparts in uniform fields.

\begin{figure*}[!t]
\centering
\includegraphics[width=0.32\textwidth]{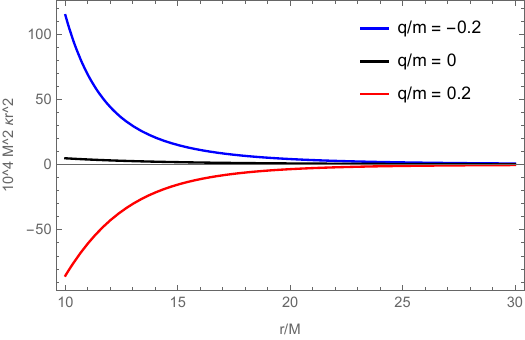}
\includegraphics[width=0.32\textwidth]{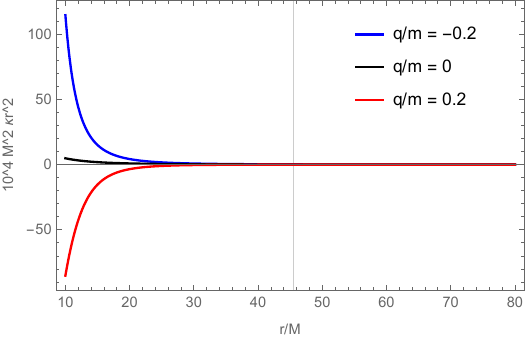}
\includegraphics[width=0.32\textwidth]{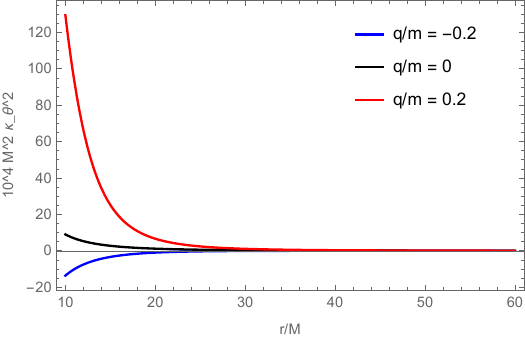}
\caption{Squared epicyclic frequencies of prograde circular orbits on the exterior branch as functions of $r/M$ for $r_0=10M$, with ${\cal B}=-0.2$ (blue), $0$ (black), and $0.2$ (red): $10^4M^2\kappa_r^2$ for $10\leq r/M\leq30$ (left) and $10\leq r/M\leq80$ (center), and $10^4M^2\kappa_\theta^2$ (right).}
\label{fig:kappa}
\end{figure*}

Figure~\ref{fig:kappa} shows the squared frequencies on the exterior branch, since a negative value signals exponential growth of the perturbation rather than an oscillation. In the left and center panels, $10^4M^2\kappa_r^2$ at $r=10M$ equals $114.67$, $4.86$, and $-85.19$ for ${\cal B}=-0.2$, $0$, and $0.2$, respectively. The neutral value agrees with Eq.~\eqref{eq:kappaRneutral}, and all curves approach zero at large $r$. The curve for ${\cal B}=-0.2$ remains positive, so these orbits are radially stable, and they have a much stronger radial restoring force than the neutral orbits. The curve for ${\cal B}=0.2$ remains negative throughout the left panel and in the center panel up to $r\simeq45.46M$, consistent with the unstable gap and with the exterior edge in Fig.~\ref{fig:Risco}. The faint vertical line in the center panel lies close to this radius. At $r=30M$, the values are $0.84$, $0.30$, and $-0.34$, which demonstrates that the coupling can still reverse the sign of $\kappa_r^2$ at three loop radii.

The right panel exhibits the opposite ordering. At $r=10M$, $10^4M^2\kappa_\theta^2$ equals $-13.64$, $9.00$, and $129.43$ for ${\cal B}=-0.2$, $0$, and $0.2$, respectively. This reversal can be read directly from Eq.~\eqref{eq:kappaTh}, which is proportional to $y\,(y+2{\cal B}\psi/r)/r^2$. On prograde orbits with $y>0$, positive ${\cal B}\psi$ reinforces the vertical restoring term, whereas negative ${\cal B}\psi$ opposes it and makes $\kappa_\theta^2$ negative once ${\cal L}_{\rm tot}+{\cal B}\psi<0$. In that case, the prograde orbits are vertically unstable even when radially stable. Solving $\kappa_\theta^2=0$ gives the outer limit of this instability, which lies at $r\simeq22.65M$ for ${\cal B}=-0.1$ and $r\simeq35.58M$ for ${\cal B}=-0.2$; the region begins just outside the loop for ${\cal B}\lesssim-0.0277$. For ${\cal B}>0$, by contrast, we find $\kappa_\theta^2>0$ throughout both radially stable bands on the sampled couplings $0<{\cal B}\leq0.2$.

Together, the panels show that the sign of the coupling exchanges radial and vertical stability on the exterior branch at fixed $r$: positive coupling near the loop produces a radial instability with vertical stability, and negative coupling produces the reverse. Therefore, a single-condition ISCO analysis is incomplete here, and radial and vertical stability must be checked together. These figures start at $r_0$ and are subject to the accuracy limits of Sec.~\ref{sec:validity}, so precise values near the loop depend on the projected source. At the same time, the qualitative exchange follows from the signs of ${\cal B}\psi$ and $\psi'$. Whether such motion is relevant to quasi-periodic oscillations would require an emission model and a treatment of the frequencies measured by a distant observer, which go beyond the present analysis.

\section{Conclusion}

In this work, we have studied the magnetic field of a stationary current loop around a static BBMB black hole and its effects on electrically charged test particles, keeping the loop at $r_0=10M$. Our main results can be summarized as follows.

We obtained exact closed-form functions for the retained dipole mode, along with elementary radial functions for all higher multipoles. The double root of the metric function eliminates the logarithmic term known in the Schwarzschild case. Every horizon-regular mode vanishes at $r=M$, and consequently the loop's flux through the degenerate horizon is zero. The exterior field approaches the Newtonian dipole form, and its direction ratio takes the simple form $r\tan\theta/(2r-M)$. Comparison with the multipole sum shows that the equatorial potential and polar field are accurate to within $10\%$ for $r\lesssim0.42\,r_0$ and $r\gtrsim3.02\,r_0$, whereas the errors near the loop reach about $40\%$ and $80\%$.

For neutral particles, the BBMB black hole has $r_{\rm ph}=2M$, $r_{\rm mb}\simeq2.618M$, and $r_{\rm ISCO}=4M$, all of which are smaller than their Schwarzschild counterparts, together with an ISCO binding-energy proxy of $8.14\%$ compared with $5.72\%$ for Schwarzschild. For charged particles, the reversal of $\psi'$ across the loop makes the dynamics depend on the region. Above ${\cal B}_c\simeq0.0119$, the prograde radially stable orbits split into interior and exterior bands separated by an unstable gap. At ${\cal B}=0.2$, the ISCO moves inward to $2.97M$, the exterior band starts at $45.46M$, and the corresponding proxies are $23.39\%$ and $0.70\%$. For negative coupling, the ISCO proxy decreases and vanishes at ${\cal B}\simeq-0.0528$.

The epicyclic analysis gives a closed-form neutral radial frequency and shows that the sign of the coupling exchanges radial and vertical stability on the exterior branch near the loop. In particular, for ${\cal B}=-0.2$, exterior orbits out to about $35.58M$ are radially stable but vertically unstable.

These results have clear limitations. The near-loop values and the critical coupling ${\cal B}_c$ belong to the projected dipole source. A calculation with higher multipoles preserves the large-coupling trends, but it moves the outer stability boundary by about $38\%$ at ${\cal B}=0.02$ and does not support the small inward ISCO shift at that coupling. We treat the BBMB background as fixed, despite its known instability, and we have not modeled the current source, radiation, or non-equatorial motion. Applying these orbital results to observed luminosities or quasi-periodic oscillations would require an accretion and radiation model. Within these limits, the loop in BBMB spacetime offers an analytically tractable case in which a degenerate horizon and a localized field jointly determine charged-particle stability.

\bibliographystyle{apsrev4-1}
\bibliography{Refs}

\end{document}